\documentclass[conference]{IEEEtran}
\IEEEoverridecommandlockouts
\usepackage{cite}
\usepackage{amsmath,amssymb,amsfonts}
\usepackage{graphicx}
\usepackage{textcomp}
\usepackage{xcolor}
\graphicspath{{figures/}}
\def\BibTeX{{\rm B\kern-.05em{\sc i\kern-.025em b}\kern-.08em
    T\kern-.1667em\lower.7ex\hbox{E}\kern-.125emX}}
\begin{document}

\title{Streaming P300 Acquisition and Statistical Signal Validation Across Five EEG Platforms: A Hardware-Agnostic BrainFlow/LSL Pipeline}

\author{
\IEEEauthorblockN{Isabella Guan}
\IEEEauthorblockA{
\textit{Lake Washington High School}\\
Kirkland, WA, USA \\
isabellaguan2009@gmail.com}
\and
\IEEEauthorblockN{Rui Liu}
\IEEEauthorblockA{
\textit{Dept. of Computer Science}\\
\textit{Stony Brook University}\\
Stony Brook, NY, USA \\
ruiliu1@cs.stonybrook.edu}
\and
\IEEEauthorblockN{Fusheng Wang}
\IEEEauthorblockA{
\textit{Dept. of Computer Science}\\
\textit{\& Biomedical Informatics}\\
\textit{Stony Brook University}\\
Stony Brook, NY, USA \\
fusheng.wang@stonybrook.edu}
}

\maketitle

\begin{abstract}
P300 spellers offer people with severe motor impairment, such as ALS, an effective communication channel and remain one of the most established surgery-free alternatives to intracortical interfaces. Advanced language models have made spellers faster and more robust, yet the hardware beneath them is under-studied. We present a hardware-agnostic, real-time P300 acquisition pipeline built on BrainFlow and Lab Streaming Layer (LSL) that runs unchanged across consumer- and research-grade EEG headsets, with permutation tests of signal separability. Using a standard 6 x 6 row/column paradigm, we piloted five configurations: a custom dry system, a custom wet/gel system, Emotiv Flex, Emotiv EPOC X, and Muse 2. The custom systems and EPOC X showed weak or inconsistent signal separability, Muse 2 had the highest acquisition reliability despite limited centro-parietal coverage, and Flex showed the most promising signal. In 20 further Flex sessions varying subject, timing, and phrase length (131 target characters), a peak-amplitude permutation test and a cross-validated xDAWN decoder both detected a significant target response under two channel-exclusion policies, with decoder AUC reaching about 0.72 after 15 repetitions. Character accuracy depended heavily on evaluation methodology: in-sample majority voting reached 94.7\%, whereas character-held-out accuracy was 31.3\% with evidence accumulated across repetitions, about three times that of held-out majority voting. These analyses indicate that Flex captured a detectable, if still weak, P300 under the studied conditions, while broader participant-level validation and improved decoding remain necessary.
\end{abstract}


\section{Introduction}

The P300 speller enables communication without muscle movement by detecting a positive event-related potential elicited around 300 ms after a subject attends to a flashing target in a row/column grid \cite{b1}, making it a promising augmentative channel for individuals with severe motor impairment, such as Amyotrophic lateral sclerosis (ALS), brainstem stroke, for whom eye movement may not provide reliable access.

Intracortical BCIs achieve high communication rates \cite{b12} but require neurosurgery unsuitable for many candidates. Non-invasive, scalp-EEG approaches such as the P300 speller remain the most established surgery-free alternative, with a growing literature on making them faster and more robust: dynamic stopping \cite{b10}, P300/eye-tracking hybrids \cite{b11}, and LLM-assisted next-character prediction \cite{b8}, reviewed in \cite{b7}.

What is comparatively under-studied is the hardware layer beneath these paradigm-level improvements: which consumer- and research-grade headsets can actually deliver a usable P300 signal, and how acquisition reliability and target-related responses vary across recording configurations and sessions. Most demonstrations validate a single, fixed headset, even though consumer-grade EEG hardware varies widely in fidelity relative to medical-grade systems \cite{b13}, and same-paradigm comparisons across multiple headsets through a shared pipeline remain uncommon -- leaving budget-limited labs unable to tell a paradigm's limitations from a given headset's.

This paper describes a hardware-agnostic streaming acquisition pipeline and a permutation-test framework for evaluating P300 signal separability across platforms, with per-session quality checks identifying acquisition failures before platform-level aggregation. Based on the pilot comparison (Section~\ref{sec:platform}), we conducted a larger Flex study with variation across sessions and subjects (Section~\ref{sec:flex20}). We also compare three character-decoding strategies on the same 20-session pool to examine differences between in-sample and held-out accuracy and the effect of combining evidence across repetitions (Section~\ref{sec:repvote}).

The paper is guided by three research questions:
\begin{itemize}
\item What signal quality and acquisition reliability are observed for each headset under a common pilot protocol?
\item Once a promising platform is identified from a small pilot, does its signal remain detectable across additional subjects, sessions, and phrase lengths?
\item How much does the resulting character-decoding accuracy depend on evaluation methodology and on how repetition-level evidence is combined?
\end{itemize}

\section{Related Work}

\subsection{The P300 Speller Paradigm and its Decoders}

The row/column P300 speller was introduced by Farwell and Donchin \cite{b1}; an alternative geometry reducing adjacency-distraction errors was proposed in \cite{b14} (Section~\ref{sec:adjacency}). Standard decoders include xDAWN spatial filtering \cite{b2}, Riemannian-geometry classification \cite{b3}, and compact CNNs such as EEGNet \cite{b4}. Benchmarking these decoders on our multi-platform dataset is future work once each platform's signal is statistically validated.

\subsection{Speed and Robustness Improvements}

Prior work targets throughput and error rate rather than hardware: dynamic-stopping methods end a trial once accumulated evidence crosses a confidence threshold \cite{b10}, P300/eye-tracking hybrids disambiguate weak trials \cite{b11}, code- or frequency-modulated stimuli increase throughput \cite{b9}, and LLM-based prediction reduces the candidate set before stimulation \cite{b8}. None of these approaches address the physical hardware question this paper focuses on.

\subsection{Streaming Infrastructure and Hardware Comparison}

Our pipeline uses the open-source BrainFlow and LSL frameworks for device abstraction and stream synchronization, respectively \cite{b5,b6}. A prior comparison of an Emotiv headset with a gel-prepared EEG system found no overall performance difference but a strong modality-by-session interaction, with Emotiv performance declining across repeated sessions \cite{b13}.

Lee et al. \cite{b15} evaluated five EEG devices using phantom-replayed signals from a public dataset, including P300 waveforms. Their analysis assessed waveform similarity using correlation coefficients, root mean square error, and ANOVA. Our study examines recordings from participants performing a row/column copy-spelling task across five configurations using a shared BrainFlow/LSL acquisition workflow, followed by a 20-session Flex study of signal separability and character decoding.

\section{System and Methods}
\label{sec:system}

\subsection{Streaming Architecture}

\begin{figure}[tb]
\centering
\includegraphics[width=\columnwidth]{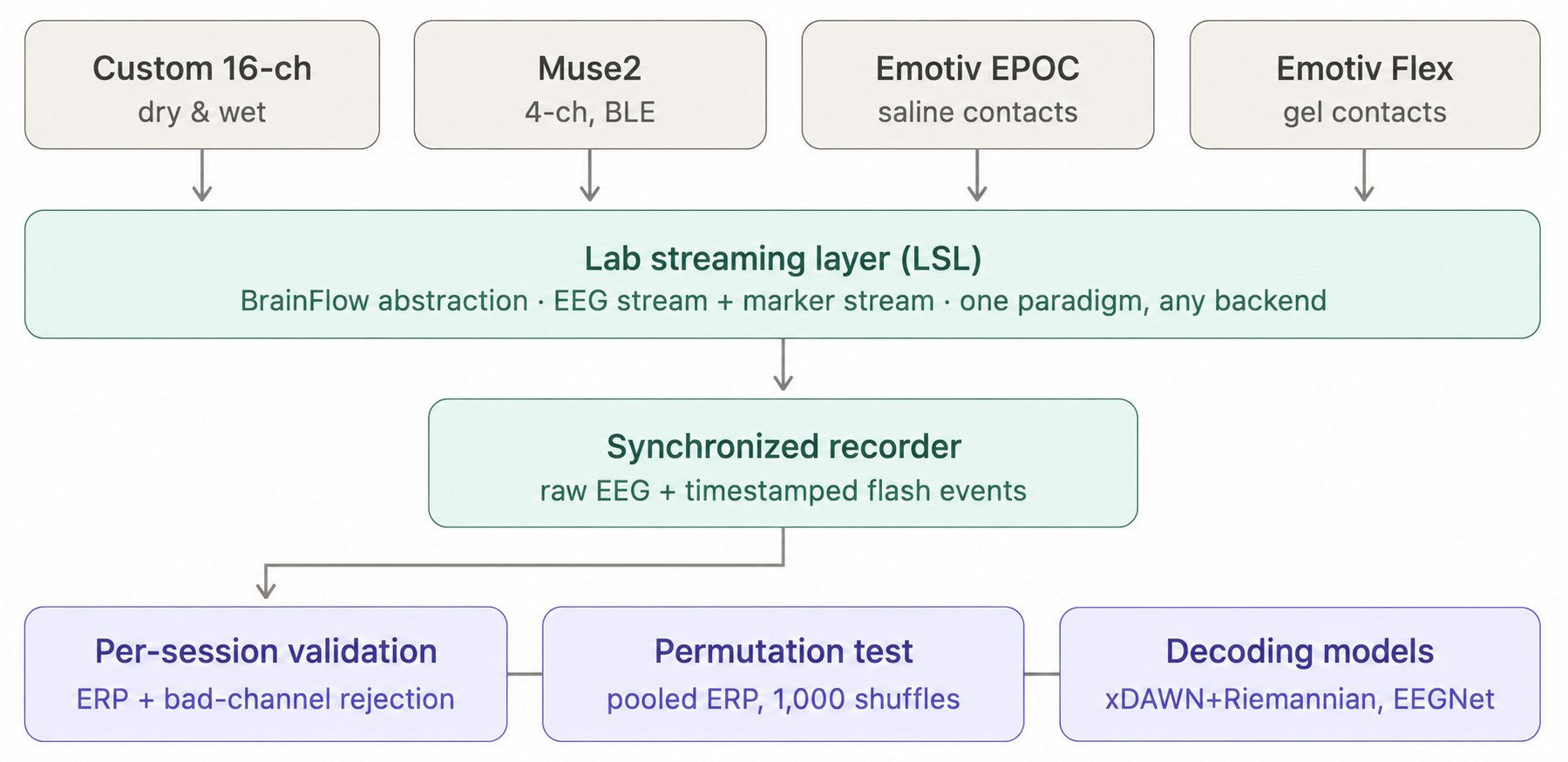}
\caption{Shared BrainFlow/LSL pipeline for synchronized EEG and stimulus-marker recording, session-level quality checks, signal validation, and character decoding.}
\label{fig:pipeline}
\end{figure}

The acquisition pipeline uses BrainFlow as a device abstraction layer, unifying acquisition across all five platforms behind a single API, and LSL as the synchronization layer, time-aligning an EEG stream with a stimulus marker stream (Fig.~\ref{fig:pipeline}). Following the LSL model, the device-facing program is a \emph{producer} publishing a named, typed stream; the recorder is a separate \emph{consumer} that discovers and subscribes to it independently, so the same consumer code runs unmodified whether the producer is a real headset or a synthetic board used for pipeline validation. The EEG stream and the marker stream (which row or column flashed) share one LSL clock, making flash-locked epoching possible, with markers aligning to the nearest EEG sample within approximately 2 ms for streams sampled at 250--256 Hz; the pipeline supports record-only, real-time, and hybrid (decode live while logging) modes, running in hybrid mode here since the target application needs both live output and a labeled log for retraining.

\subsection{Paradigm}
\label{sec:paradigm}

Stimuli were presented in a 6x6 row/column flash grid (Farwell-Donchin layout): 15 repetitions per character, 100 ms flashes, a 75 ms inter-stimulus gap, and a 175 ms stimulus onset asynchrony (SOA). Each repetition flashes all 12 stimulus codes once in random order (codes 1-6 = columns, 7-12 = rows). Data collection used a copy-spelling protocol: since the target phrase is known in advance, every flash is automatically labeled target or non-target at record time, yielding fully supervised data without manual annotation. Each character trial was preceded by a 1 s cue and followed by a 1.5 s rest period.

\begin{figure}[tb]
\centering
\includegraphics[width=\columnwidth]{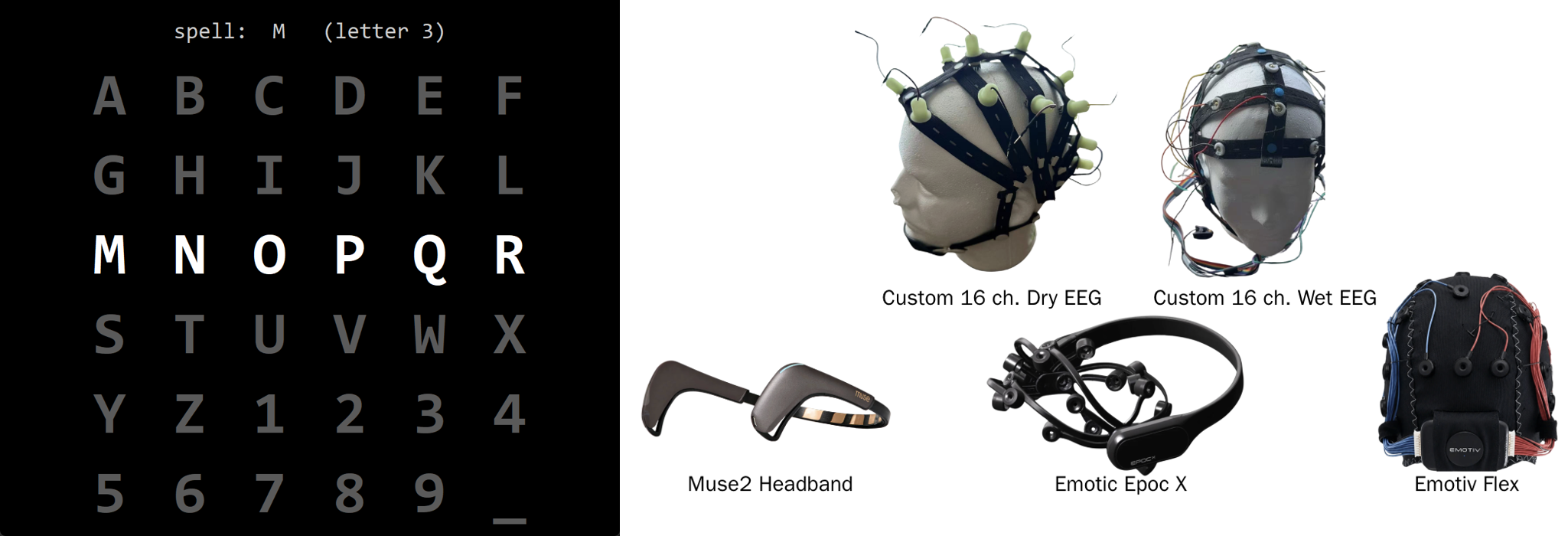}
\caption{(a) A single row-flash stimulus frame from the copy-spelling task. The current target character (``M'', letter 3 of the phrase) is shown above the grid; the row containing M-R is illuminated during this flash. (b) The five EEG platforms and configurations evaluated in this work.}
\label{fig:speller}
\end{figure}

\subsection{Paradigm Refinement: Target Flash Adjacency}
\label{sec:adjacency}

During pilot data collection we observed that a target character's row and column flashes could occasionally land back-to-back, or split across a repetition boundary -- collapsing two events that should each independently evoke a P300 into one closely-spaced pair, an instance of the row/column adjacency-distraction confound previously identified for this paradigm family \cite{b14}. For the extended 20-session study (Section~\ref{sec:flex20}) we therefore added a scheduling constraint: at least one non-target flash must separate a character's target row and column codes, guaranteeing a minimum 350 ms separation between the two target-flash onsets (versus 175 ms without it, given the 175 ms SOA) -- introduced after the two pilot Flex sessions and in place for all 20 extended-validation sessions.

\subsection{Multi-Session Pooling}
\label{sec:pooling}

A single recording session of a short phrase yields only a handful of character trials -- typically five -- too few for a label-shuffled permutation test to have meaningful power. Epochs from multiple sessions of the same platform are therefore pooled before the analyses in Section~\ref{sec:results}, with each character assigned a global identifier so characters from different sessions never collide. For decoder cross-validation, all epochs from the same character are kept in the same fold, preventing flashes from one character from appearing in both training and evaluation data. We therefore report character trials, rather than individual epochs, as the primary count of repeated observations.

\section{Results}
\label{sec:results}

\subsection{Platform Comparison (Pilot)}
\label{sec:platform}

The five platforms group into three device families, summarized in Table~\ref{tab:platforms} (methodology in Section~\ref{sec:sigval}). The custom dry configuration had 14 of 16 channels unusable because of poor scalp contact. Switching to wet/gel electrodes resolved the dead-channel issue, although intermittent glitches and DC offset remained. EPOC X and Flex were considered promising for data acquisition. Flex's 32-electrode montage provides coverage near centro-parietal Pz/Cz, but its pilot result was based on only 2 sessions; Section~\ref{sec:flex20} reports a further 20-session study. Muse 2 had the most synchronized sessions (12) and the highest observed acquisition reliability, although its montage does not cover centro-parietal sites where P300 is strongest. These observations concern different aspects of acquisition: Muse 2 had more recorded sessions, whereas Flex had the lowest exploratory p-value.

\begin{table}[htbp]
\caption{Summary of All Five Platforms (Pilot)}
\begin{center}
\begin{tabular}{|l|c|c|c|c|}
\hline
\textbf{Platform} & \textbf{Ch.} & \textbf{Sess.} & \textbf{Reliability} & \textbf{Best p} \\
\hline
Custom 16ch, dry & 16 & 2 & low & 0.833 \\
\hline
Custom 16ch, wet/gel & 16 & 4 & medium & 0.542 \\
\hline
Emotiv EPOC X & 14 & 2 & variable & 0.508 \\
\hline
Emotiv Flex & 32 & 2 & medium & \textbf{0.034} \\
\hline
Muse 2 & 4 & 12 & high & 0.084 \\
\hline
\end{tabular}
\label{tab:platforms}
\end{center}
\end{table}

\subsection{Signal Validation Methodology}
\label{sec:sigval}

We assessed target-minus-non-target ERP differences within the 250--500 ms P300 window. The magnitude and shape of these differences varied across channels.

Signal separability is validated per channel with a label-shuffled permutation test: the true target-minus-non-target peak amplitude in the P300 window is computed, then recomputed under 1,000 random label permutations to build a null distribution, with the p-value taken as the fraction of permuted peaks meeting or exceeding the observed one; Section~\ref{sec:platform} reports the best (lowest) p-value across channels for each platform. This nonparametric test suits noise characteristics that differ substantially across platforms. EEG was band-pass filtered 1--30 Hz and epoched from $-100$ to 800 ms relative to flash onset; all amplitudes are in physical microvolts (\textmu V), with each epoch baseline-corrected to its own pre-stimulus mean.

\begin{figure}[tb]
\centering
\includegraphics[width=\columnwidth]{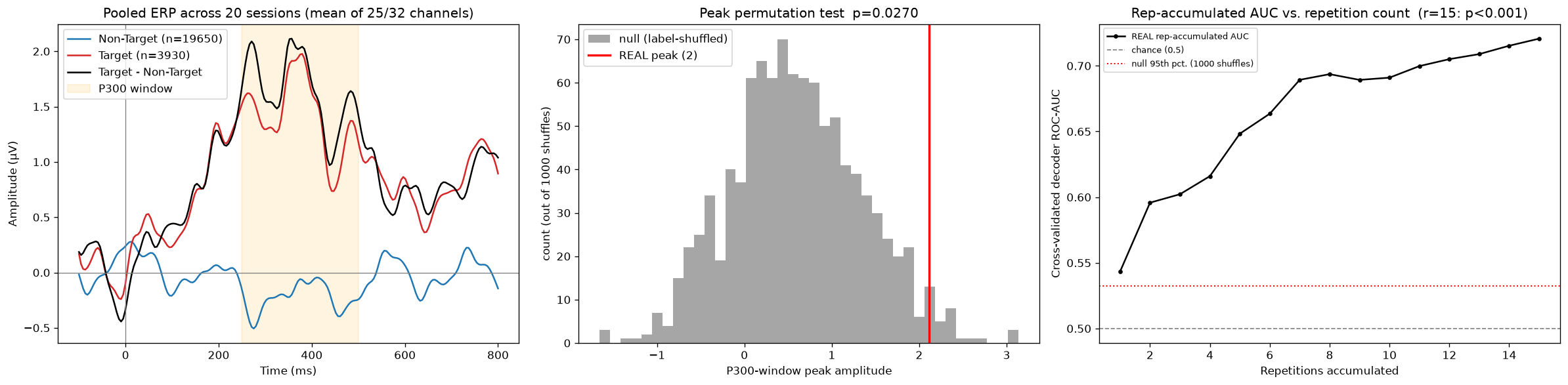}
\caption{Pooled ERP, peak-amplitude permutation test, and rep-accumulated xDAWN decoder AUC (rising with repetition count) for the $\geq$25\%-of-sessions configuration; both tests' full numbers and the any-session-policy comparison are in Table~\ref{tab:flex20}, while the 17-session peak-test sensitivity comparison is in Table~\ref{tab:sensitivity}.}
\label{fig:flex20combined}
\end{figure}

Because Section~\ref{sec:platform} reports each platform's \emph{best} p-value across all channels, this figure is the minimum of many uncorrected per-channel tests (Section~\ref{sec:threats}); we therefore treat it only as an exploratory statistic, not confirmatory evidence. Section~\ref{sec:flex20} reports a follow-up analysis that avoids this issue, illustrated in Fig.~\ref{fig:flex20combined} with real data from the 20-session Flex validation.

\subsection{Extended Validation: Emotiv Flex Across 20 Sessions}
\label{sec:flex20}

Section~\ref{sec:platform} identified Emotiv Flex as a candidate for further study, although the pilot included only two sessions and used uncorrected channel-wise tests. We collected 20 additional sessions, varying subject, timing, and phrase length (4--11 characters), all using the adjacency refinement described in Section~\ref{sec:adjacency}. Pooling these sessions yielded 23,580 epochs from 131 character trials. We compared two channel-exclusion policies: the any-session policy excluded 17 of 32 channels, including Pz, whereas the $\geq$25\%-of-sessions policy excluded 7 channels and retained Pz.

Each configuration was evaluated using a channel-averaged peak-amplitude test and an xDAWN + Riemannian + logistic-regression decoder. Five-fold cross-validation grouped by character trial produced an out-of-fold score for each flash. For each candidate stimulus code, scores were summed across repetitions 1 through $r$. Each character's target row and target column were treated as the two positive codes among the 12, with the remaining 10 treated as negative. We pooled the labels and accumulated scores across character trials to calculate a single ROC-AUC at each repetition count, yielding a curve for $r=1$--15.

The peak-amplitude test followed the label-shuffling procedure described in Section~\ref{sec:sigval}, applied to the channel-averaged response. For the decoder-AUC test, we kept the out-of-fold scores fixed and performed 1,000 target-code-shuffle permutations. Reusing the scores avoided refitting the decoder for each permutation and gave a minimum reportable p-value of approximately 0.001.

\begin{table}[htbp]
\caption{Extended Flex Validation (20 Sessions) Under Two Channel-Exclusion Policies}
\begin{center}
\setlength{\tabcolsep}{3pt}
\begin{tabular}{|l|c|c|c|c|}
\hline
\textbf{Policy} & \textbf{Ch.} & \textbf{Peak p} & \textbf{AUC ($r{=}1{\to}15$)} & \textbf{AUC p} \\
\hline
Any-session exclusion & 15/32 & 0.021 & 0.532 $\to$ 0.708 & $<0.001$ \\
\hline
$\geq$25\%-of-sessions & 25/32 & 0.027 & 0.543 $\to$ 0.721 & $<0.001$ \\
\hline
\end{tabular}
\label{tab:flex20}
\end{center}
\end{table}

Table~\ref{tab:flex20} summarizes both tests under both policies. The peak p-values were similar (0.021 vs.\ 0.027), as were the rep-accumulated AUC endpoints. AUC increased from ~0.53--0.54 at one repetition to 0.708--0.721 at 15, exceeding all 1,000 label-permuted AUCs ($p<0.001$). In the smaller pools examined earlier (2- and 4-session), only the peak test reached significance. The results therefore differed with pool size. Section~\ref{sec:repvote} evaluates character accuracy using the 20 sessions and 131 characters.

\subsection{Sensitivity Analysis: A Post Hoc Session Subset}
\label{sec:sensitivity}

Of the 20 sessions, three -- 13 (``HUNGRY'', $-0.97$ \textmu V), 14 (``HAPPY'', $-0.17$ \textmu V), and 18 (``CANDY'', $-0.13$ \textmu V) -- had a negative per-session Target-minus-Non-Target peak. All three were flagged ``low/absent'' by the session-level heuristic, and no session flagged ``P300 detected'' had a negative peak. We examined the effect of excluding these three sessions by rerunning the peak test on a post hoc 17-session subset using the same $\geq$25\%-of-sessions channel-exclusion policy.

\begin{table}[tb]
\caption{Full 20-Session Dataset vs.\ 17-Session Sensitivity Subset}
\centering
\setlength{\tabcolsep}{4pt}
\begin{tabular}{|l|c|c|c|c|}
\hline
\textbf{Dataset} & \shortstack{\textbf{Epochs}\\\textbf{(tgt/non-tgt)}} & \shortstack{\textbf{Charac-}\\\textbf{ters}} & \shortstack{\textbf{Ch.}\\\textbf{kept}} & \shortstack{\textbf{Peak}\\\textbf{p}} \\
\hline
All 20 sessions & 3{,}930/19{,}650 & 131 & 25/32 & 0.027 \\
\hline
17-session sensitivity subset & 3{,}450/17{,}250 & 115 & 26/32 & \textbf{0.015} \\
\hline
\end{tabular}
\label{tab:sensitivity}
\end{table}

The peak-test p-value decreased from 0.027 to 0.015 after removing sessions 13, 14, and 18. Because the subset was defined post hoc and the channel count changed ($25\to26$), this comparison does not establish that the excluded sessions contained only noise. The result is exploratory, not confirmatory; the full 20-session analysis in Table~\ref{tab:flex20} remains primary.

\subsection{Character Decoding: Comparing In-Sample, Out-of-Fold, and Rep-Accumulated Strategies}
\label{sec:repvote}

We next evaluated character accuracy and the effect of combining the 15 repetitions. The three strategies use the same 131 characters and the same xDAWN + Riemannian + logistic-regression architecture, but differ in training scope and decision rule. Strategies (2) and (3) share out-of-fold scores from 5-fold, character-grouped cross-validation across all 20 sessions. Strategy (1) fits a separate decoder within each session and evaluates it on the same flashes used for training:

\begin{enumerate}
\item \textbf{In-sample majority vote.} Fit once per session on all its flashes; each repetition is decoded from its own 12 flashes, and the character's letter is the majority vote across 15 guesses.
\item \textbf{Out-of-fold majority vote.} Same logic as (1), but scores come from Section~\ref{sec:flex20}'s character-grouped, 5-fold cross-validated decoder: a character's own flashes never enter the fold that scores it.
\item \textbf{Out-of-fold accumulated evidence.} Same scores as (2), but summed across all 15 repetitions before picking the highest-scoring row and column -- Section~\ref{sec:flex20}'s accumulation logic, now scored as an exact match rather than a ranking statistic.
\end{enumerate}

\begin{table}[tb]
\caption{Strategy (1) In-Sample Majority-Vote Decoding, Per Session (20 Sessions)}
\centering
\scriptsize
\setlength{\tabcolsep}{3pt}
\begin{tabular}{|c|l|l|c|}
\hline
\textbf{Sess.} & \textbf{Phrase} & \textbf{Winning-vote count / char (of 15)} & \textbf{Correct} \\
\hline
1 & CSSRE & 5/5/2/8/6 & 3/5 \\
\hline
2 & CSIRE & 8/7/6/5/9 & 5/5 \\
\hline
3 & HELLO\_WORLD & 2/3/3/5/3/5/4/7/1/3/2 & 8/11 \\
\hline
4 & EYE\_CAN\_DO & 4/3/3/4/7/5/5/5/5/3 & 9/10 \\
\hline
5 & GOOD\_FOCUS & 9/10/14/13/4/6/11/11/6/5 & 10/10 \\
\hline
6 & LOVELY\_DAY & 5/11/8/6/5/3/5/4/6/5 & 9/10 \\
\hline
7 & NEED\_WATER & 11/7/12/8/7/11/9/11/8/6 & 10/10 \\
\hline
8 & OPEN\_WINDOW & 10/12/12/10/10/13/14/13/12/12/8 & 11/11 \\
\hline
9 & WATER & 13/11/11/7/11 & 5/5 \\
\hline
10 & CLOTH & 6/5/7/9/5 & 5/5 \\
\hline
11 & STAND & 10/10/10/11/11 & 5/5 \\
\hline
12 & SLEPT & 14/13/13/13/14 & 5/5 \\
\hline
13 & HUNGRY & 10/12/10/9/11/9 & 6/6 \\
\hline
14 & HAPPY & 13/10/11/14/9 & 5/5 \\
\hline
15 & PEACH & 15/13/13/13/15 & 5/5 \\
\hline
16 & MUSIC & 7/6/5/6/7 & 5/5 \\
\hline
17 & WORSE & 9/10/10/8/7 & 5/5 \\
\hline
18 & CANDY & 11/9/14/10/12 & 5/5 \\
\hline
19 & RICE & 14/13/15/13 & 4/4 \\
\hline
20 & NICE & 6/9/11/8 & 4/4 \\
\hline
\textbf{All 20} & --- & --- & \textbf{124/131} \\
\hline
\end{tabular}
\label{tab:repvote}
\end{table}

Across all 131 characters, strategy (1) reached 94.7\% (124/131), strategy (2) reached 10.7\% (14/131), and strategy (3) reached 31.3\% (41/131). Strategy (1) fits xDAWN's spatial filters and the downstream logistic regression on the same labeled flashes it later scores, including all 15 repetitions of each character. Its accuracy therefore does not estimate performance on unseen trials. It also differs from strategies (2) and (3) in training scope, so the full difference between 94.7\% and 10.7--31.3\% cannot be attributed to evaluation leakage alone. Out-of-fold single-repetition accuracy (strategy 3, $r=1$) was 2.3\%, close to the nominal 2.78\% chance level. Strategy (1) did not show near-chance accuracy at any repetition count, but these were training-set results. Even under strategy (1), individual repetitions were often split widely: for one session-3 character (target ``E''), the fifteen per-repetition guesses spanned ten distinct letters (K, M, 9, M, Z, C, A, P, E, V, A, 3, 9, E, E), with ``E'' winning only 3-to-2 over the nearest runner-up -- illustrating why the paradigm accumulates evidence across repetitions rather than deciding from one.

Strategies (2) and (3) use identical out-of-fold scores, allowing a direct comparison of their decision rules. Accumulating continuous evidence yielded 31.3\% accuracy, approximately three times the 10.7\% obtained by repetition-level voting. Voting reduces each repetition to a single character choice, whereas accumulation retains the continuous scores across all 15 repetitions before selecting a row and column (Section~\ref{sec:paradigm}). Accumulation performed better in this dataset, although the accuracy remains too low for practical spelling.

\section{Discussion}
\label{sec:threats}

Under pilot conditions, Muse 2 had the highest observed acquisition consistency despite limited coverage; Emotiv Flex had the lowest exploratory peak-test p-value but required gel. Limitations of this study are that setup time, comfort, cost, and caregiver burden were not measured, so these observations do not support a deployment recommendation yet. The custom dry-electrode board needs better scalp contact before reevaluation. The uncorrected best-across-channels statistic (Section~\ref{sec:platform}) also gives platforms with more channels more opportunities to obtain a low p-value. Section~\ref{sec:flex20} uses a channel-averaged peak test and a multivariate decoder under two exclusion policies, while the other platforms remain at pilot scale (2-12 sessions). Character-held-out evaluation keeps each character trial's flashes together but does not hold out entire sessions or participants. Performance on new sessions and participants therefore remains unestablished. The Flex results support further  pipeline evaluation, but the current character accuracy and limited validation across participants leave substantial work before practical use. Future work will prioritize:

\begin{itemize}
\item \textbf{More training data and broader validation}, across more participants under prespecified quality criteria, followed by an evaluation of dynamic stopping to balance accuracy and selection time.
\item \textbf{Decoder comparison}: benchmarking EEGNet \cite{b4} against xDAWN+Riemannian, including information transfer rate \cite{b2,b3,b4}.
\item \textbf{Language models beyond the character level}: an LLM predicting likely words/sentences from accumulated evidence \cite{b8}, rather than decoding characters independently.
\end{itemize}

\section{Conclusion}
\label{sec:conclusion}

We implemented a BrainFlow/LSL streaming pipeline supporting five EEG platforms under a shared P300 paradigm, with permutation tests for evaluating signal separability. The main findings were:

\begin{itemize}
\item \textbf{Cross-platform acquisition.} The comparison used live streaming data; Lee et al. \cite{b15} used phantom-replayed waveforms. Muse 2 had the highest observed acquisition reliability across 12 sessions, with an exploratory peak-test p-value of 0.084. Flex had the lowest pilot p-value (p=0.034), based on two sessions.

\item \textbf{Pooled signal analysis.} Across 20 sessions (131 characters), channel-averaged peak amplitude and rep-accumulated, cross-validated xDAWN decoder AUC both exceeded the test thresholds, whereas only the peak test did in smaller pools.

\item \textbf{Character decoding.} Using the same out-of-fold scores, accumulation across 15 repetitions yielded 31.3\% character accuracy, compared with 10.7\% for majority voting. Separate in-sample evaluation yielded 94.7\%, but this training-set result does not estimate performance on unseen trials. Future evaluations should choose the grouping level to match the intended generalization setting.

\item \textbf{Sensitivity analyses and stimulus scheduling.} Both tests exceeded their thresholds under two channel-exclusion policies. The extended study separated target row/column flashes by $\geq$350 ms. Excluding three sessions in a post hoc analysis yielded a peak-test p-value of 0.015; the full 20-session analysis remains primary.
\end{itemize}
Together, these results show that Flex captures a detectable but weak P300 under the studied conditions, and that reported spelling accuracy depends heavily on the evaluation protocol and on how repetition-level evidence is combined. The shared pipeline and two-test validation framework provide a reusable basis for comparing headsets under one paradigm before P300 decoder development further.

\section*{ACKNOWLEDGMENT}
The authors thank Dr. Fusheng Wang and the CSIRE program at Stony Brook University for supporting this research.


\end{document}